%% file: ijcai22.tex
\documentclass{article}
\usepackage{ijcai22}
\usepackage{tikz}
\def\checkmark{\tikz\fill[scale=0.3](0,.35) -- (.25,0) -- (1,.7) -- (.25,.15) -- cycle;} 
\usepackage{microtype}
\usepackage{graphics}
\usepackage{bbm}
\usepackage{caption}
\usepackage{algorithm}
\usepackage{algorithmic}

\usepackage{graphicx}
\usepackage{multirow}
\usepackage{subfigure}
\usepackage{tikz}

\usepackage{times}
\usepackage{soul}
\usepackage{url}
\usepackage[hidelinks]{hyperref}
\usepackage[utf8]{inputenc}
\usepackage{graphicx}
\usepackage{amsmath}
\usepackage{amsthm}
\usepackage{booktabs}
\usepackage{algorithm}
\usepackage{algorithmic}
\newcommand{\ignore}[1]{}
\usepackage{bbold}

\usepackage{bbm}
\usepackage{bbold}
\usepackage{bbding}
\usepackage{dsfont}

\usepackage{amsmath}
\usepackage{bm}

\title{Unsupervised Context Aware Sentence Representation Pretraining for Multi-lingual Dense Retrieval}

\author{
Ning Wu$^1$
\and
Yaobo Liang$^2$\and
Houxing Ren$^1$\and
Linjun Shou$^1$\and
Nan Duan$^2$\and
Ming Gong$^{1}$\And
Daxin Jiang$^1$
\affiliations
$^1$Microsoft STCA\\
$^2$Microsoft Research Asia
\emails
\{wuning, yalia, v-houxingren, lisho, nanduan, migon, djiang\}@microsoft.com
}

\begin{document}

\maketitle

\begin{abstract}

\input{2022_naacl_draft/abstract}

\end{abstract}

\input{2022_naacl_draft/intro}

\input{2022_naacl_draft/related}

\input{2022_naacl_draft/unicoder}

\input{2022_naacl_draft/experiments}

\input{2022_naacl_draft/conclusion}

\bibliographystyle{named}
\bibliography{naacl2022}

\end{document}

%% file: 2022_naacl_draft/abstract.tex


Recent research demonstrates the effectiveness of using pretrained language models (PLM) to improve dense retrieval and multilingual dense retrieval. In this work, we present a simple but effective monolingual pretraining task called contrastive context prediction~(CCP) to learn sentence representation by modeling sentence level contextual relation. By pushing the embedding of sentences in a local context closer and pushing random negative samples away, different languages could form isomorphic structure, then sentence pairs in two different languages will be automatically aligned.  Our experiments show that model collapse and information leakage are very easy to happen during contrastive training of language model, but language-specific memory bank and asymmetric batch normalization operation play an essential role in preventing collapsing and information leakage, respectively. Besides, a post-processing for sentence embedding is also very effective to achieve better retrieval performance.  On the multilingual sentence retrieval task Tatoeba, our model achieves new SOTA results among methods without using bilingual data. Our model also shows larger gain on Tatoeba when transferring between non-English pairs. On two multi-lingual query-passage retrieval tasks, XOR Retrieve and Mr.TYDI, our model even achieves two SOTA results in both zero-shot and supervised setting among all pretraining models using bilingual data. The pretrained model and code are available in this link:\url{https://github.com/wuning0929/CCP_IJCAI22}.

%% file: 2022_naacl_draft/intro.tex
\section{Introduction}
\label{sec:intro}

Nowadays, cross-lingual pre-training~\cite{devlin-etal-2019-bert,conneau2019cross,huang2019unicoder,xlmr} has achieved great performance on cross-lingual transfer learning tasks. These pre-trained models could fine-tune on one language and directly test on other languages. Cross-lingual sentence representation models like  InfoXLM~\cite{chi2020infoxlm} and LaBSE~\cite{feng2020language} target to generate good cross-lingual representation without fine-tuning. These models use contrastive loss to make bilingual sentence pairs have similar embeddings and achieve great performance on bilingual sentence retrieval tasks.



However, there are two potential problems for these methods. First, most of them rely on bilingual corpus, which are not always available, especially for low-resource languages and non-English languages pairs. Only using English related bilingual pairs will limit the transferability between non-English pairs.
Second, multilingual dense retrieval tasks such as XOR Retrieve and Mr.TYDI require the model to map semantic related query and passage to similar position in embedding space, but existed methods only could map bilingual sentence pairs with same meaning to similar embedding.
 




\begin{figure}[h]
\centering
\subfigure[ XLM-R.]{
\begin{minipage}[t]{0.47\linewidth}
\centering
\includegraphics[width=1.0\linewidth]{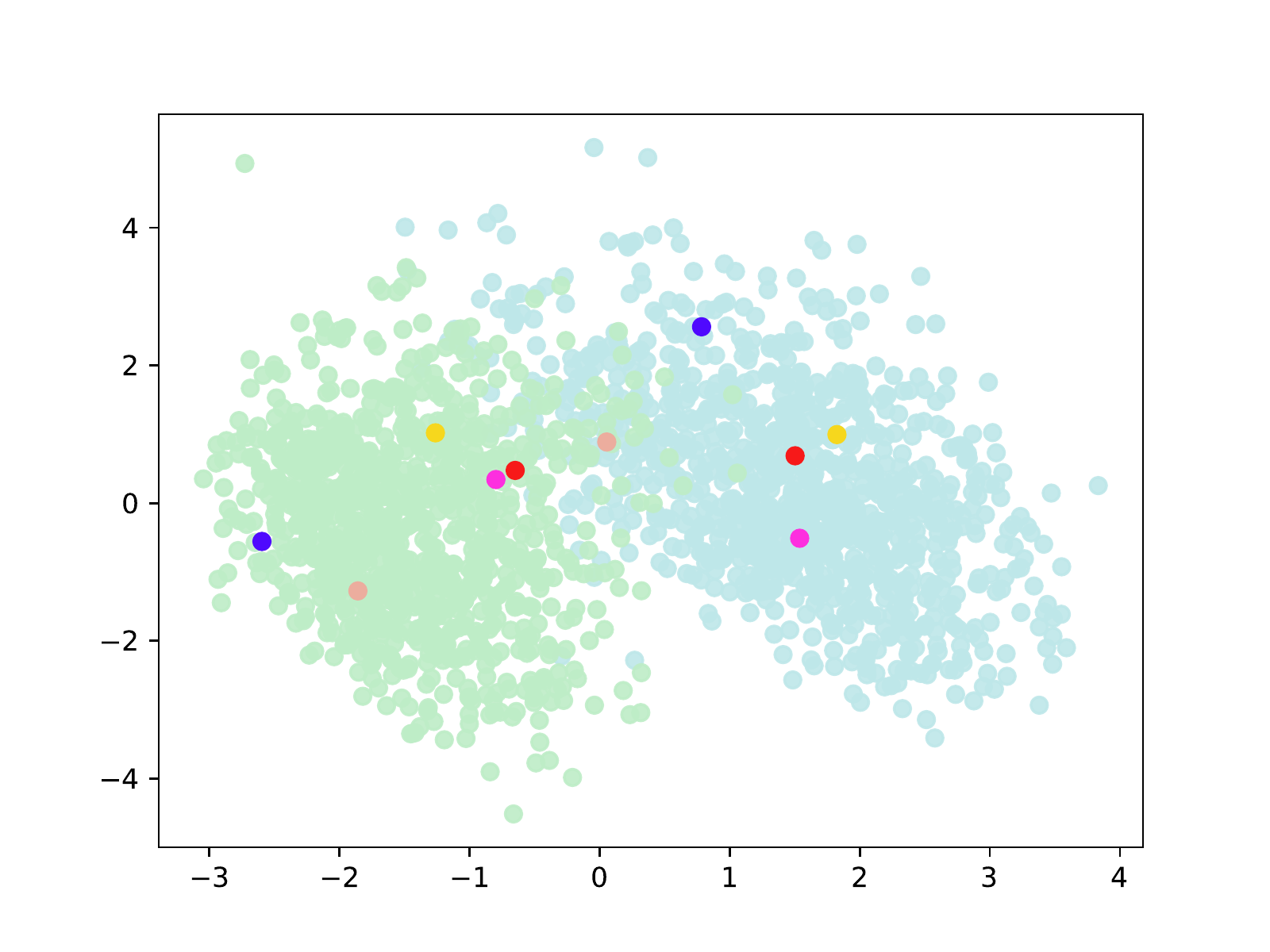}
\end{minipage}\label{fig:case-study-a}

}%
\subfigure[ XLM-R with Calibration.]{
\begin{minipage}[t]{0.48\linewidth}
\centering
\includegraphics[width=1.0\linewidth]{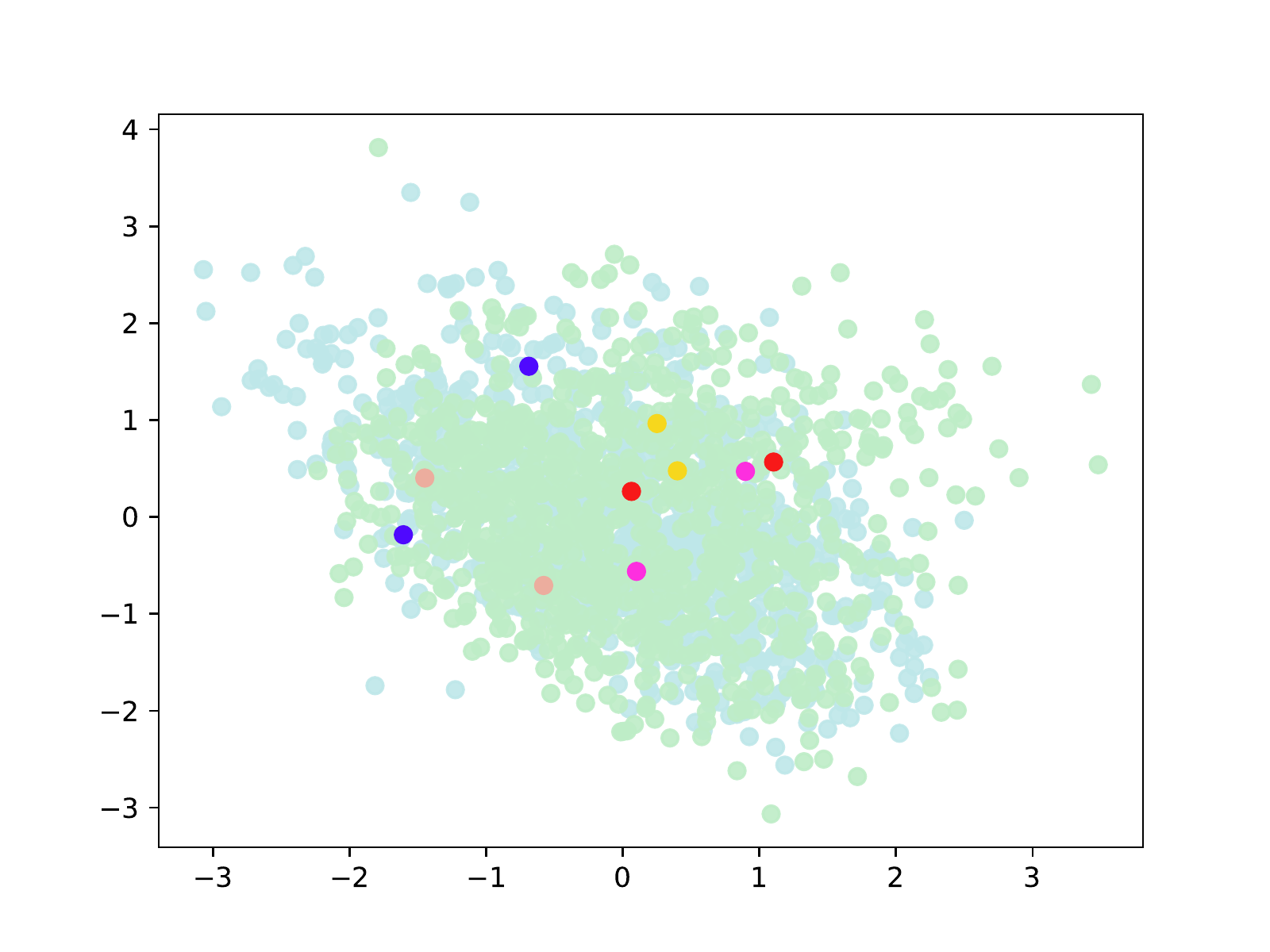}
\end{minipage}\label{fig:case-study-b}
}%
\vskip\baselineskip
\subfigure[ CCP without Calibration.]{
\begin{minipage}[t]{0.48\linewidth}
\centering
\includegraphics[width=1.0\linewidth]{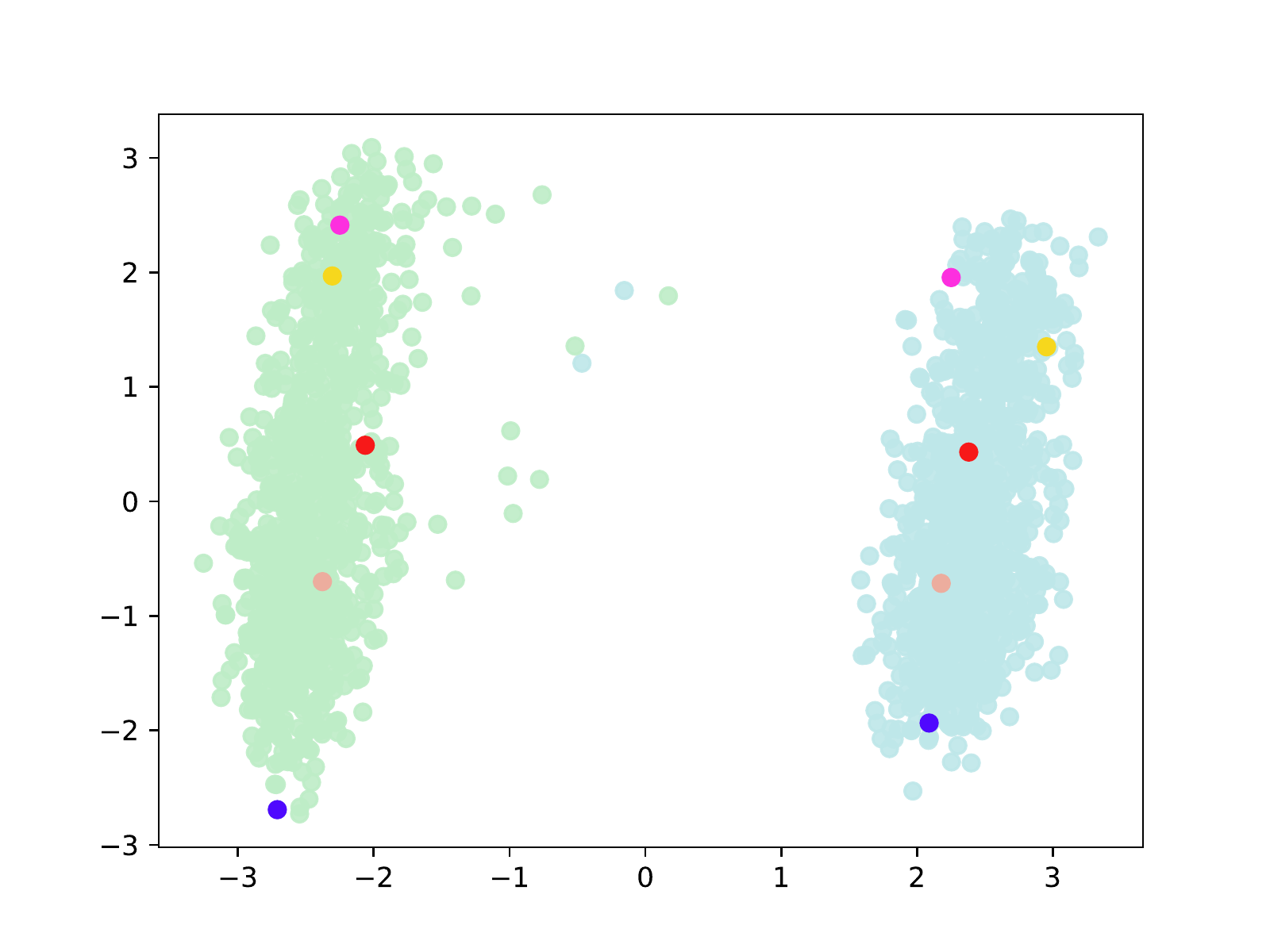}
\end{minipage}\label{fig:case-study-c}
}%
\subfigure[CCP with Calibration.]{
\begin{minipage}[t]{0.48\linewidth}
\centering
\includegraphics[width=1.0\linewidth]{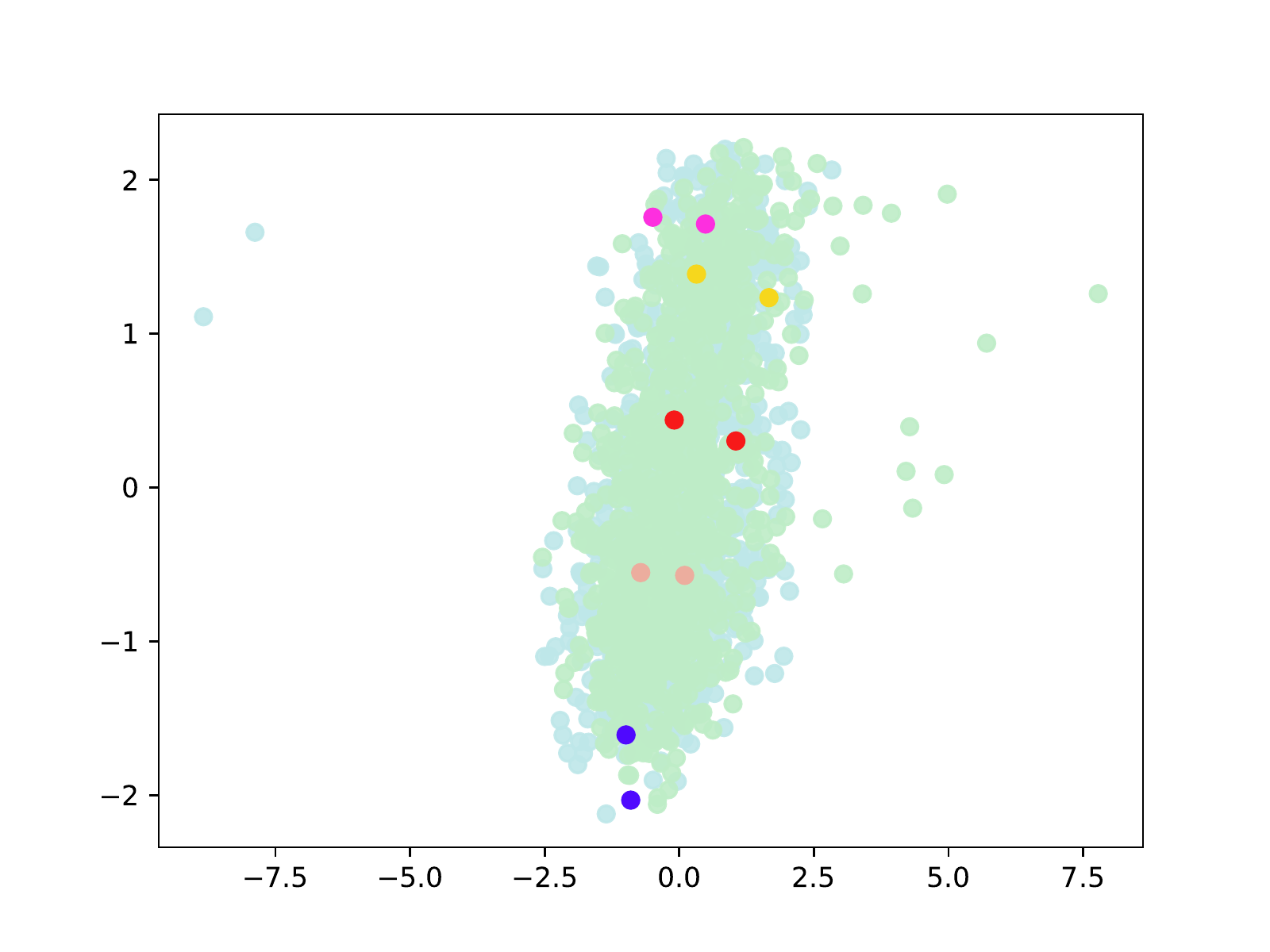}
\end{minipage}\label{fig:case-study-d}
}%
\centering
\caption{Visualization of the representation output of four models. A thousand of bilingual sentence pairs from Tatoeba are converted into representations by four models. The dimension of these representations is reduced from 768 to 2 by Principle Component Analysis~(PCA), so that a sentence can be mapped into a point on figure (a), (b), (c) and (d). For each figure, we highlight five bilingual pairs, and the two highlighted cycles in the same color denote a bilingual sentence pair from English and Arabic, respectively. } \label{fig:case-study}
\end{figure}



\ignore{
Inspired by the success of contrastive learning in computer vision. We want to explore whether it could help models generate better cross-lingual sentence representation. 
}

Inspired by recent progress of contrastive learning on dense retrieve, we propose a new pretraining task called Contrastive Context Preiction~(CCP). CCP targets on constructing isomorphic embedding space by modeling the sentence level contextual relation in long documents. Formally, a document is a sequence of sentences. For each center sentence, we define the sentence in the window centered on it as a context sentence. First, CCP will encode each sentence into a vector. Given embedding of center sentence $s_c$, the model need to select correct context sentence $s$ out of thousands random sampled sentences and vice versa. With contrastive context prediction, we could estimate the mutual information $I(s|s_c)$ of contextual relation. Our experiments show that the embeddings of CCP have isomorphic structure across different languages. Furthermore, we do cross-lingual calibration to further improve the alignment. We illustrate our ideas with Figure~\ref{fig:case-study}, which visualizes the English and Arabic embedding space from XLM-R and our models. We randomly highlight five points for each languages and the sentences with same meaning have same color. The distribution of two languages shown in Figure~\ref{fig:case-study-a} doesn't have an obvious pattern. With contrastive context prediction, the five points show similar relative position and the shape of all points are similar to each other, but the embeddings of two languages are spread in two regions of latent space. After cross-lingual calibration, the points with same color almost have similar position.



We evaluate our model on bilingual sentence retrieval task Tatoeba, which can test whether our model could generate similar embeddings for two sentences with same meaning but from different languages. Our model achieves SOTA results among methods without bilingual data, and our results are very close to the model with bilingual data. Besides, our model shows better cross-lingual transferability between non-English languages pairs.
On two multi-lingual query-passage retrieval tasks, XOR Retrieve and Mr.TYDI, CCP achieves new SOTA results among all pretraining models using bilingual data.


Our contribution can be summarized as:
\begin{itemize}
\item We propose a contrastive context prediction pretraining~(CCP) task that is capable of learning isomorphic representations for each language without parallel data. Our model can achieve excellent performance on multi-lingual retrieval, especially between two non English-centric languages-pairs.

\item We design an effective contrastive pretraining framework for sentence embedding pretraining. It consists of language-specific memory bank and projection head with asymmetric batch normalization. Both of them play essential role in preventing collapsing. And we also observe the offset phenomenon on on bilingual sentence representation pair produced by our model. To align the sentence embedding space between different languages, we propose cross-lingual calibration to align the bilingual sentence pair into the same position in latent space.

\item We conduct evaluation experiments upon three types of multi-lingual retrieval tasks. Extensive results on the three datasets have shown superiority of the proposed model in both effectiveness and rubustness.

\end{itemize}



%% file: 2022_naacl_draft/related.tex
\section{Related Work}
\label{sec:rel}

\paragraph{Cross-lingual Pre-trained Model} Multilingual BERT (M-BERT)~\cite{devlin-etal-2019-bert} performs pre-training based on the multilingual corpus with the masked language model task. By sharing the model parameters and the vocabulary across all languages, M-BERT obtains the cross-lingual capability over 102 languages. XLM~\cite{conneau2019cross} performs cross-lingual pre-training based on multilingual corpus and bilingual corpus, by introducing the translation language model task into pre-training. Based on XLM, Unicoder~\cite{huang2019unicoder} uses more cross-lingual pre-training tasks and achieves better results on XNLI. XLM-R~\cite{xlmr} is a RoBERTa~\cite{liu2019roberta}-version XLM without using translation language model in pre-training. It is trained based on a much larger multilingual corpus (i.e. Common Crawl) and becomes the new state-of-the-art.  
Both these models and our model could achieve good performance after fine-tuning. Our model also could produce good cross-lingual sentence embedding without fine-tuning.

\paragraph{Dense Passage Retrieval} 

\cite{lee-etal-2019-latent} proposed a simple Inverse Cloze Task
(ICT) method to further continue-train BERT. REALM\cite{guu2020realm} is an end-to end co-training framework for reader and retriever. \cite{karpukhin-etal-2020-dense} is the first to discover that careful fine-tuning can learn effective dense retriever directly from BERT. Later works then started to investigate ways to further improve
fine-tuning. ANCE ~\cite{xiong2020approximate} selects hard training negatives globally from the entire corpus, using an asynchronously updated ANN index.
\cite{qu-etal-2021-rocketqa} proposed the RocketQA fine-tuning pipeline which hugely advanced
the performance of dense retrievers. coCondenser~\cite{gao2021unsupervised} is one of the best dense passage retrieval model on MS-MARCO, Natural Question  and Trivia QA. It adds an unsupervised corpus-level contrastive loss to warm up the passage embedding space.




\paragraph{Contrastive Learning} 
CPC~\cite{oord2018representation} predicts the future in latent space by using powerful auto-regressive models. It uses a probabilistic contrastive loss which induces the latent space to capture information that is maximally useful to predict future samples.
\cite{wu2018unsupervised} presents an unsupervised feature learning approach by maximizing distinction between instances via a novel non-parametric softmax formulation, which is so-called memory bank mechanism.
SimCLR~\cite{chen2020simple} simplifies recently proposed contrastive self-supervised learning algorithms without requiring specialized architectures or a memory bank. 
MoCo~\cite{he2020momentum} maintains a dynamic dictionaries by implementing a momentum-based moving average mechanism of the query encoder. 
InfoXLM~\cite{chi2020infoxlm} contains a pre-training task based on contrastive learning. Given a bilingual sentence pair, they regard them as two views of the same meaning, and encourage their encoded representations to be more similar than the negative examples. LaBSE\cite{feng2020language} combines masked language model~(MLM) and translation language model~(TLM)\cite{conneau2019cross} pretraining with a
translation ranking task using bi-directional dual encoders. CLEAR ~\cite{wu2020clear} performs contrastive learning on for sentence representation. It employs multiple sentence-level augmentation strategies in order to learn a noise-invariant sentence representation. We follow SimCLR on the detailed implementation of contrastive learning. Different from these works, we leverage contrastive learning to model the sentence-level contextual relation in natural languages.

%% file: 2022_naacl_draft/unicoder.tex

\section{Methodology}

\begin{figure}[t]
\begin{center}
 \includegraphics[width=0.9\linewidth]{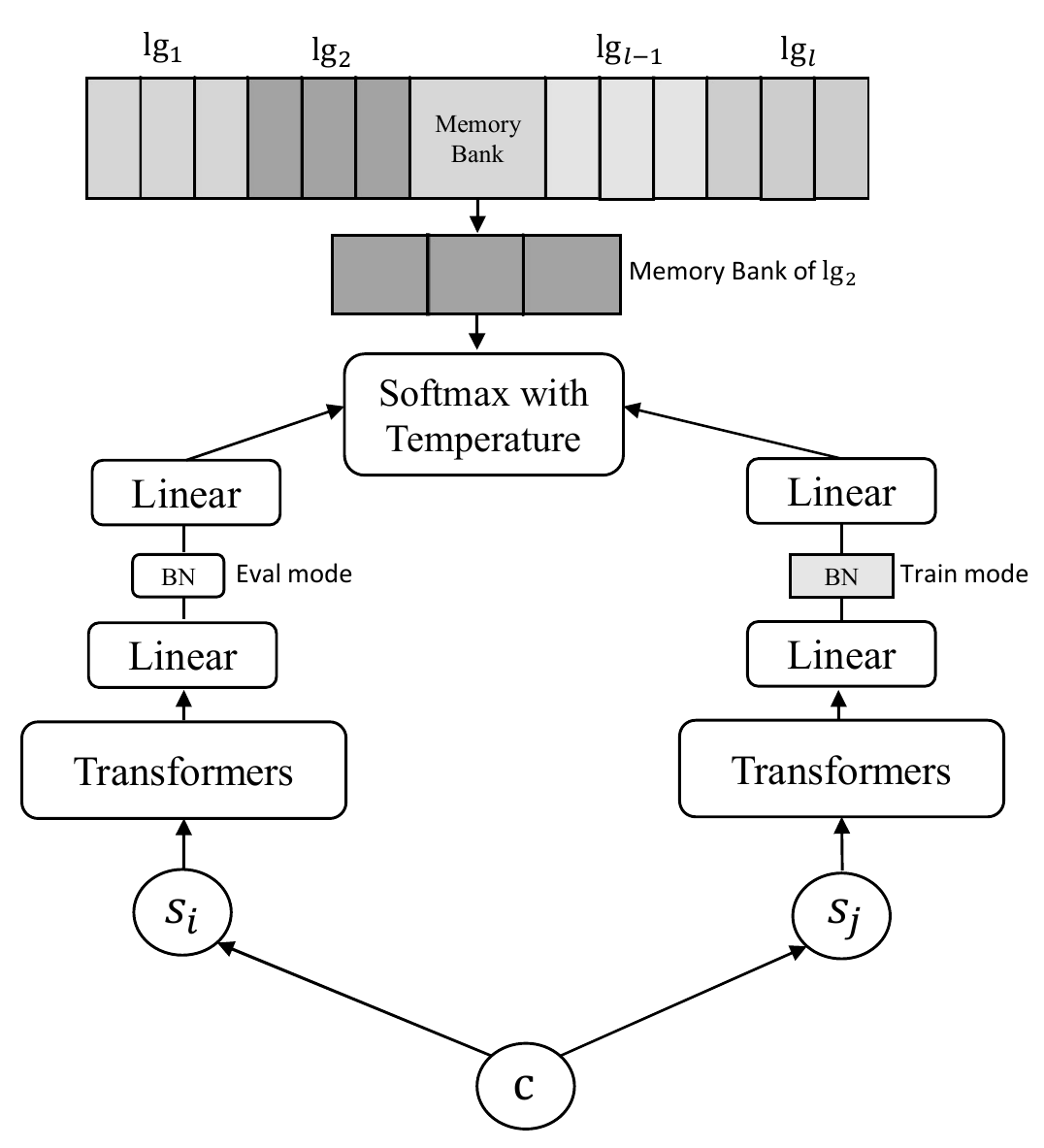}\vspace{-.3cm}
\caption{Overview of our framework of contrastive context prediction, the proposed representation learning framework. Two contextual sentences are encoded by training mode batch normalization and testing mode batch normalization, respectively. Finally, we add negative samples of the same language from memory bank in contrastive loss calculation. }
 \label{fig-illustration}\vspace{-.3cm}
\end{center}
\end{figure}



Our method contains two steps. First, we train the model with Masked Language Model and Contrastive Context Prediction. This step targets on building isotropic sentence embedding space for each language. The learned sentence embeddings from different languages will have good cross-lingual property after they became isomorphic by CCP task. Just as Figure~\ref{fig:case-study-c} shows, different languages are spread in different region of latent space with similar structure. Thus in second step, we use cross-lingual calibration to further align different languages.


\subsection{Contrastive Loss for Context Prediction}
Our method targets to model the sentence-level contextual relation. Formally, a document $D$ is a sequence of sentences $(s_1, s_2, ..., s_l)$. For each center sentence $s_c$, we define its contextual set as $Context(s_c)=\{s_{p} | c-w<=p<=c+w, p\ne c\}$, $w$ is radius of window which represents the maximum distance between center sentence and context sentence. Our model will model the relation between center sentence and its contextual sentences. In this subsection, we introduce the details of contrastive loss $\mathcal{L}_N(s_i, s_c)$.

\paragraph{Scoring Function} The scoring function $f(s_i,s_c)$ takes two sentences as input and output a score. To begin with, we encode $s$ and $s_c$ to a vector separately. In this step, we choose a Transformer-based encoder. We use the corresponding representation of a manually-inserted token [CLS] as the embedding of the whole sentence. Then, we add a non-linear neural network, namely Projection Head, to further map the vector to a new space. The projection head consists of two linear layers and one batch normalization between two layers. We denote the representation of $s_i$ and $s_c$ as $z_i$ and $z_c$ respectively. Following SimCLR~\cite{chen2020simple}, we only use it when computing contrastive loss and abandon it after pre-training. Projection Head could help model learn general representations and will not overfit to contrastive loss. Finally, we choose scoring function as $f(s_i, s_c)=exp(cosine(z_i, z_c)/\tau)$, where $\tau$ represents temperature and is a hyper-parameter.

\paragraph{Language-specific Memory Bank} In \cite{oord2018representation}, the lower bound of contrastive loss becomes tighter as negative samples number $N$ becomes larger. To further increase $N$ while batch size is limited by GPU memory, we use memory bank to store the embeddings from recent batches and use them in the training of current batch. Because we are handling multiple languages at same time, we tried two strategies:  plain memory bank and language-specific memory bank, and find later one is better. The "language-specific" means that our memory bank will tag the embeddings by language. For each language, it will only use the embeddings from the same language in training. We tried to use shared memory bank for all languages. Since a shared memory bank provides negative samples from various languages, the model will focus on classifying the language of sentences, rather than learning contextual relation. Hence the loss could be very small because language classification is very easy, but the cross-lingual performance will be very poor.
The memory bank is maintained in FIFO~(First-In-First-Out) manner. During each iteration, the representation $z_i$ as well as the network parameters $\theta$ are optimized via Adam. Then $z_i$ is added to $M$, and the oldest representation in the memory bank is deleted.


\paragraph{Asymmetric Batch Normalization} 
Batch normalization is an essential part of projection head. However, traditional batch normalization is easy to be trapped in information leak problem~\cite{he2020momentum}, which means that the contrastive loss is very small but the evaluation results on down stream tasks are very low. Hence, we propose asymmetric batch normalization to avoid information leak. It is more efficient than shuffle batch normalization~\cite{he2020momentum}, which requires communication between GPUs. In training procedure, the mode of batch normalization will be changed in training mode and testing mode alternately and the mode of batch normalization in two projection heads should be kept different in training. In testing mode, the batch normalization will use running mean and running variance to replace batch mean and batch variance, which is able to prevent information leak. The detail training procedure can be seen in Algorithm ~\ref{alg-training}. Compared with shuffle batch normalization proposed in~\cite{he2020momentum}, our method is very easy to implement and has good performance.

\begin{algorithm}[t]\small
\caption{The training algorithm for the contrastive context prediction task.}\label{alg-training}
\begin{algorithmic}[1]
\STATE \textbf{Input:} Batch size $N$, constant $\tau$, structure of $f$, $g$. 
\STATE \textbf{Output:} Model parameters $\Theta$.
\FOR{ sampled minibatch $\{\bm{c}_k\}^{N}_{W}$}
\STATE draw center sentence $s_c$ from context $c_w$.
\STATE randomly draw target sentence $s_i$ from context $c_w$. 
\STATE $\bm{h}_c$=$f(s_c)$; $\bm{h}_i$=$f(s_i)$;
\IF{training mode $e=0$}
    \STATE $g_{c}()=g().\text{train}()$; $g_{i}()=g().\text{eval}()$;
    \STATE $\bm{z}_c$=$g_{c}(\bm{h}_c)$; $\bm{z}_i$=$g_{i}(\bm{h}_i)$; $e=1$;
\ELSE
    \STATE $g_{c}()=g().\text{eval}()$; $g_{i}()=g().\text{train}()$;
    \STATE $\bm{z}_c$=$g_{c}(\bm{h}_c)$; $\bm{z}_i$=$g_{i}(\bm{h}_i)$; $e=0$;
\ENDIF

\STATE \begin{small}$l^w_{c,i}=-\log \frac{\exp(\cos(z_c,z_i)/\tau)}{\sum_{k=1}^{2N+M_{lg(i)}} \mathds{1}_{[k \ne c]} \exp(\cos (z_c,z_k)/\tau) }$\end{small}

\STATE $\mathcal{L}_{CL}=\sum_{c=1}^{2N} \sum_{i=1}^{2N}   m(s_c,s_i)l_{c,i}^w$
\STATE $m(s_c,s_i)=1$ means $c$ and $i$ exists in the same local window, unless $m(s_c,s_i)=0$.

\STATE update networks $f$ and $g$ to minimize $\mathcal{L}_{CL}$.
\ENDFOR
\RETURN encoder network $f(\cdot)$, and throw away $g(\cdot)$ 
\end{algorithmic}
\end{algorithm}

\ignore{
\paratitle{Contrastive Loss}
A contrastive learning loss function defined for a contrastive prediction task, i.e., trying to predict positive augmentation pair ($\widetilde{s}_1$, $\widetilde{s}_2$) in the set {$\widetilde{s}$}. For example, two sentence in the same context form the positive pair, while all other instances from the same mini-batch and memory bank are regarded as negative samples for them.  The contrastive learning loss has been tremendously used in previous work. The loss for a positive pair is defined as 

\begin{equation}
\resizebox{.9\hsize}{!}{
l_{c,i}=-\log \frac{\exp(\cos(z_c,z_i)/\tau)}{\sum_{k=1}^{2N+M_{lg(c)}} \mathbbm{1}_{[k \ne c]} \exp(\cos (z_c,z_i)/\tau) }  
}
\end{equation}

where $\mathbbm{1}_{[k \ne c]}$ is the indicator function to judge whether $k \ne c$, $\tau$ is a temperature parameter, $N$ is the size of mini-batch, $M$ is the size of memory bank.  $\cos(u,v)=u^Tv/( \left\|u\right\|_2 \left\|v\right\|_2)$ denotes the cosine similarity of two vectors $u$ and $v$. The overall contrastive learning loss is defined as the sum of all positive pairs' loss in a mini-batch:

\begin{equation}
\mathcal{L}_{CL}= \sum_{i=1}^{2N} \sum_{j=1}^{2N} m(s_i,s_j)l_{i,j}  
\end{equation}
where m(i,j) is a function returns 1 when i and j is a positive pair, returns 0 otherwise.

\subsection{The Combined Loss for Pre-training}
For the purpose of grabbing both token-level and sentence-level features, we use a combined loss of MLM objective and ST objective to get the overall loss:

\begin{equation}
\mathcal{L}_{total}= \mathcal{L}_{MLM} + \mathcal{L}_{ST} 
\end{equation}

where $\mathcal{L}_{MLM}$ is calculated through predicting the random-masked tokens in set {s} as described in RoBERTa. Our pre-training task is to minimize the $\mathcal{L}_{total}$.
}

\subsection{Cross-lingual Calibration}
After we acquire isomorphic sentence embeddings by pretraining, in order to better align between languages, we further propose cross-lingual calibration for sentence representation. The cross-lingual calibration consists of three operations: shifting, scaling and rotating. We do three operations separately to better understand the properties of latent space. For shifting, we compute the mean of sentence embedding $\bm{\mu}$ from different languages. Then we acquire shifted mean sentence representation by subtracting corresponding language mean vector. For Scaling, we compute the variance of sentence embedding $\bm{\sigma}$ for all languages based on Common Crawler corpus. For a shifted sentence, we acquire a scaled sentence representation by dividing corresponding language variance vector. Finally, we learn the rotation matrix in unsupervised method proposed by \cite{conneau2017word}. With orthogonal rotation matrix $\bm{W}_{i,j}$ , sentence embedding of $lg(i)$ can be mapped to $lg(j)$. We provide detailed description of this part in the supplementary materials.

\ignore{
\begin{itemize}
\item \textbf{Shifting:}
After pre-training, we compute the mean of sentence embedding $\bm{\mu}$ from different languages. Then we acquire shifted mean sentence representation by subtracting corresponding language mean vector. $lg(i)$ denotes language identifier of sentence $s_i$.

\begin{equation}
\bm{s}_i^{1}= \bm{s}_i-\bm{\mu}_{lg(i)} 
\end{equation}

\item \textbf{Scaling:}
Similarly, we compute the variance of sentence embedding $\bm{\sigma}$ for all languages based on Common Crawler corpus. For a shifted sentence, we acquire a scaled sentence representation by dividing corresponding language variance vector.

\begin{equation}
\bm{s}_i^{2}= \frac{\bm{s}^1_i}{\bm{\sigma}_{lg(i)}} \end{equation}

\item \textbf{Rotating:}
To align bilingual sentence embedding pair, the final step is rotating. With orthogonal matrix $\bm{W}_{i,j}$, sentence embedding of $lg(i)$ can be mapped to $lg(j)$:

\begin{equation}
\bm{s}_i^{3}= \bm{s}_i^{2} \bm{W}_{i,j} 
\end{equation}

To learn the rotation matrix $\bm{W}_{i,j}$, we collect sentence representations from language $lg(i)$ and language $lg(j)$, then we learn the rotation matrix in unsupervised method proposed by \cite{conneau2017word}.

\end{itemize}
}

\ignore{
Similarly, we statistic the variance of sentence embedding $\sigma_{lg(i)}$ for all languages based on Common Crawler corpus. For a sentence

In this section, we present a 
simple but effective method which is able to improve various down-streaming tasks.

 then normalize the sentence embedding into normal distribution by subtract corresponding language mean and divide variance.
For sentence representation $\bm{z}_i$, we can acquire a new representation $\hat{\bm{z}}_i$ after the proposed normalization.
\begin{equation}
\hat{\bm{z}}_i=\frac{\bm{z}_i - \bm{\mu}_{lg(i)}}{\sigma_{lg(i)}}  \nonumber
\end{equation}

In our experiments, we found that the language normalization can significantly improve down-streaming tasks on models which don't use bilingual data, while for these models which utilize bilingual pair, they can't get so much improvement from the language normalization.
}

%% file: 2022_naacl_draft/experiments.tex
\section{Experiments}
\label{sec:experiments}
In this section, we first set up the experiments, and then present the performance comparison and result analysis.

\begin{table*}
\small
    \begin{center}
          \resizebox{1\linewidth}{!}{
    	\begin{tabular}{ccccccccccccccccccc}
        	 \hline
         Type & Methods	 & FR & ES & DE & EL & BG & RU & TR & AR & VI & TH & ZH & HI & SW & UR & AVG & AVG$_{all}$ \\
            \hline

            		\multirow{2}{*}{M} & XLM-R  &
            		56.0 & 56.7 & 72.0 & 35.4 & 48.5 & 50.0 & 49.6 & 31.5 & 45.9 & 36.5 & 42.1 & 47.8 & 11.0 & 32.0 & 43.9 & -\\

& CRISS & 92.7 & 96.3 & 98.0 & - & - & 90.3 & 92.9 & 78.0 & 92.8 & - & 85.6 & 92.2 & - & - & - & - \\
\hline
            		\multirow{4}{*}{B+M} & 
 INFOXLM & 83.7 & 87.8 & 94.7 & 67.1 & 78.9 & 84.9 & 83.5 & 63.5 & 89.8 & 86.7 & 84.9 & 86.4 & 35.8 & 69.4 & 78.4 & -\\
& LASER & 95.7 & 98.0 & 97.3 & 95.0 & 95.1 & 94.6 & 97.6 & 91.9 & 96.8 & 95.4 & 95.5 & 94.7 & 57.6 & 81.9 & 91.9 & 65.5 \\
& Unicoder & 81.6 & 86.5 & 93.8 & 67.2 & 77.5 & 81.6 & 76.7 & 53.4 & 80.9 & 70.2 & 87.7 & 73.8 & 30.3 & 59.2 & 72.9 & -\\
& LABSE & \textbf{96.0} & \textbf{98.4} & \textbf{99.4} & \textbf{96.6} & \textbf{95.7} & \textbf{95.3} & \textbf{98.4} & \textbf{91.0} & \textbf{97.8} & \textbf{97.1} & \textbf{96.2}  & \textbf{97.7}  & \textbf{88.6}  & \textbf{95.4}  & \textbf{96.0} & \textbf{83.7} \\
\hline
            		\multirow{2}{*}{M} & 
 CCP & 93.8 & 96.6 & 98.5 & 87.6 & 88.2 & 92.0 & 95.2 & 81.5 & 94.3 & 90.2 & 91.8 & 90.4 & 50.5 & 82.8 & 88.1 & -\\
& CCP+Calibration & 94.9 & 97.2 & 99.0 & 93.0 & 90.3 & 93.5 & 97.1 & 87.9 & 96.3 & 95.3 & 95.0 & 96.2 & 64.2 & 91.3 & 92.2 & 78.8\\
    \hline
\end{tabular}
        }
        \end{center}
    \caption{Evaluation results on English-centric cross-lingual sentence retrieval. Type means if a model uses bilingual data~(B) and monolingual~(M) data in pre-training.  Given each model, the corresponding retrieval results on all languages are listed in the same row. We report the average Top-1 accuracy of two direction(e.g. EN-FR and FR-EN). AVG$_{all}$ is the average of 112 languages that Tatoeba supports. }\label{tab:en2xx}
\end{table*}

\begin{table*}
\small
    \begin{center}
          \resizebox{1\linewidth}{!}{
    	\begin{tabular}{cccccccccccccccccc}
        	\hline

          Type & Methods	& DE-EL
 & DE-IT & RU-NL & FR-DE & IT-RU & AR-RU & ZH-ES & ZH-JA & JA-FR & ES-PT & IT-RO & SV-DA & DA-NO & NL-DE & UR-RU & AVG  \\
         	\hline
        	\multirow{2}{*}{M} &
XLM-R & 54.6 & 54.9 & 66.6 & 75.8 & 47.4 & 43.4 & 47.0 & 52.8 & 42.5 & 76.0 & 51.4 & 81.1 & 89.4 & 66.7 & 85.6 & 62.3\\
&CRISS & - & 77.5 & 83.5 & 88.4 & 74.7 & 79.6 & 80.7 & 72.5 & 74.0 & - & - & - & - &- & - & -\\
\hline
\multirow{2}{*}{B+M} &
Unicoder & 68.5 & 68.5 & 77.3 & 81.9 &  64.7 &  64.5 & 65.9 & 67.3 &  55.3 & 82.4 & 61.9 & 85.4 & 91.9 & 72.4 &  90.1 &  73.2\\
&LABSE & \textbf{85.6} & \textbf{80.3} & \textbf{89.9} & \textbf{90.4} & \textbf{78.1} & \textbf{87.3} & \textbf{88.5} & \textbf{88.6} & \textbf{89.1} & \textbf{84.3} & \textbf{72.6} & 87.6 & \textbf{93.5} & 78.1 & 91.7 &  \textbf{85.7}\\
\hline
\multirow{2}{*}{M} &
CCP & 83.5 & 78.6 & 87.5 & 89.0 & 76.3 & 83.6 & 83.9 & 84.1 & 82.6 & 84.0 & 70.7 & 87.7& 93.1 & 78.2 & 92.1 & 83.7\\
&CCP+Calibration & 85.2 & 80.0 & 89.1 & 89.7 & 77.8 & 86.2 & 87.4 & 86.5 & 87.6 & 83.7 & \textbf{71.2} & \textbf{87.9
}& 93.2 & \textbf{78.4} & \textbf{92.3} & \textbf{85.1}\\
        	\hline
        \end{tabular}
        }
        \end{center}
    \caption{Evaluation results on Non-English cross-lingual sentence retrieval. Type means if a model uses bilingual data~(B) and monolingual~(M) data in pre-training. Given each model, the corresponding retrieval results on all languages are listed in the same row. We report the average Top-1 accuracy of two direction(e.g. DE-EL and EL-DE).}\label{tab:xx2xx}
\end{table*}

\ignore{
\begin{table*}
\small
    \begin{center}
          \resizebox{1\linewidth}{!}{
    	\begin{tabular}{cccccccccccccccccc}
        	\hline

          Methods & Metrics	& DE
 & ES & EN & FR & ID & IT & PT & & RU & ZH & AVG  \\
        	\hline

\multirow{2}{*}{CCP} &
MRR@10 & 0.173 & 0.180 & 0.229 & 0.161 & 0.158 & 0.132 & 0.173  & 0.109 & 0.092 & 0.156\\
&Recall@1000 & 0.805 & 0.825 & 0.911 & 0.792 & 0.776  & 0.724 & 0.812 & 0.665 &  0.594 & 0.767\\
        	\hline
        	
\multirow{2}{*}{InfoXLM} &
MRR@10 & 0.152 & 0.154 & 0.203 & 0.140 & 0.124 & 0.116 & 0.148  & 0.095 & 0.081 & 0.135\\
&Recall@1000 & 0.797 & 0.817 & 0.904 & 0.787 & 0.742  & 0.713 & 0.797 & 0.664 &  0.586 & 0.756\\
        	\hline

\multirow{2}{*}{CRISS} &
MRR@10 & - & - & - & - & - & - & - & - & - & -\\
&Recall@1000 & - & - & - & - & - & - & - & - & - & - \\
        	\hline

\multirow{2}{*}{XLMR} &
MRR@10 & - & - & - & - & - & - & - & - & - & -\\
&Recall@1000 & - & - & - & - & - & - & - & - & - & - \\
        	\hline

        \end{tabular}
        }
        \end{center}
    \caption{Evaluation results on m-MSMARCO.}\label{tab:xx2xx}
\end{table*}
\subsection{Pre-training Details}
}

Our CCP model has 1024 hidden units, 16 attention heads and 24 layers in encoder. Following \cite{wenzek2019ccnet}, we collect a clean version of Common Crawl as pre-training corpus. It leads to 2,500GB multilingual corpus covering 108 languages. We first initialize the CCP model with XLM-R\cite{xlmr}, and then run continued pre-training with the accumulated 2,048 batch size with gradient accumulation and a memory bank of 32768. One 2,048 batch consists of many small batches whose size is 32 for CCP.  We use Adam Optimizer with a linear warm-up and set the learning rate to 3e-5. We select two pre-training tasks randomly in different batches. This costs 7 days on 16 V100 for CCP model.

\subsection{Baselines}
Here are the baselines for our experiments.

\ \textbullet \textit{M-BERT}~\cite{devlin-etal-2019-bert} is a multilingual version of BERT.

\textbullet \textit{XLM-R}~\cite{xlmr} uses a Transformer-based masked language model on one hundred languages.

\textbullet \textit{InfoXLM}~\cite{chi2020infoxlm} formulates a cross-lingual pre-training as maximizing mutual information between multilingual multi-granularity text. 

\textbullet \textit{Unicoder}~\cite{liang-etal-2020-xglue} uses mask language model and translation language model as pre-training tasks.

\textbullet \textit{CRISS}~\cite{tran2020cross} utilizes cross-lingual retrieval for iterative training.

\textbullet \textit{LaBSE}~\cite{feng2020language} formulates a translation ranking task using bi-directional dual encoders.


We present the comparison between our method and ABSent~\cite{fu2020absent} in appendix, since it only reports their performance on 3 language pairs of tatoeba,  All transformer models in this paper use Bert-large structure except CRISS and LaBSE. CRISS follows mBART structure with 24 layers transformer, and LaBSE follows Bert-base structure with a 500k size vocabulary, which is twice as large as our model.



\subsection{Cross-lingual Sentence Retrieval}
To better evaluate the performance on massive languages, we adopt the Tatoeba corpus introduced by \cite{artetxe2019massively}. It consists of 1,000 English-centric sentence pairs for 112 languages and the task aims to find the nearest neighbor for each sentence in the other language using cosine similarity distance. To compare with previous model, we only report results on 14 language in experiments, and we present all results in Table 2 of supplementary material.  Besides the English-centric dataset constructed by \cite{artetxe2019massively}, we choose 14 language-pairs which don't contain English in the Tatoeba raw dataset, and we present results on 50 language-pairs which don't contain English in Table 2 of supplementary material. Following \cite{artetxe2019massively}, we extract 1000 sentence-pairs for each language pair and test the pre-trained model on the Tatoeba dataset without fine-tuning directly. The accuracy for each language pair is computed. We report CCP in Table~\ref{tab:en2xx} and Table~\ref{tab:xx2xx}. For large models, we use the the averaged hidden vectors in the 14-th layers as sentence representation for sentence retrieval.

\ignore{
We test the pre-trained model on Tatoeba without fine-tune. We use the averaged hidden vectors in the middle layers to retrieve bilingual sentence pair.   We consider two setting: English-pivot and Other-pivot. English-pivot sentence retrieval means that use English to retrieve other language and use other language to retrieve English, This setting has been tremendously used in previous work, because most cross-lingual model use English as pivot language in training phrase. The second setting means that we use two other language to retrieve each other. This setting is usually ignored by previous work. We construct retrieval dataset from Tatoeba. For the first setting, we choose 14 language-pair which occurred in XNLI. For the second setting,   Accuracy of the sentence retrieval is used as the metric.
}
In Table~\ref{tab:en2xx}, (1) we find CCP performs significantly better than XLM-R and CRISS and achieves new SOTA results among methods without using bilingual data on the Tatoeba dataset. (2) Compared with models using bilingual data, CCP performs better than InfoXLM~\cite{chi2020infoxlm}, Unicoder~\cite{liang-etal-2020-xglue} and  LASER~\cite{artetxe2019massively}, but it is worse than LaBSE~\cite{feng2020language}, which is the SOTA model with bilingual data on Tatoeba. LaBSE uses bidirectional dual encoders with 8192 batch size to learn cross-lingual sentence representation. The translation ranking task is similar to contrastive learning task, which is severely affected by the batch size. Limited by hardware, we can only perform contrastive learning with a batch size of 32 , it's hopeful our model will have a better result with larger batch.


In Table~\ref{tab:xx2xx}, we find (1) CCP performs significantly better than XLM-R and CRISS. We observe that the performance of CRISS decreases obviously, because CRISS only mines English-centric bilingual pairs and it is easy to overfit English-centric sentence retrieval. (2) For models using bilingual corpus, CCP performs better than Unicoder and its performance is very close to LaBSE. Since LaBSE is trained on English-centric bilingual corpus, its performance decreases severely on sentence retrieval between two non-English sentences, which means it is overfitting on English-centric sentence retrieval. Compared with these English-centric models~(LABSE, Unicoder, InfoXLM, CRISS) , our model is  more general, and it does not rely on English data in down-streaming tasks. So our model has lesser performance loss when it is evaluated on non English-centric sentence retrieval. 

These two experiments proved that CCP has good cross-lingual sentence retrieval performance both on English-centric and Non-English languages pairs. Our performance is better than all models without using bilingual data and slightly worse than LaBSE which used bilingual data.


\ignore{
Furthermore, to explore the cross-lingual understanding ability of our model, we try to use our model to measure the relation between different language. In table 3, we choose 9 language from three families: Romance, Germanic and Slavic. It's obvious that language from the same family has higher retrieval score. For example, the accuracy of sentence retrieval between Spanish~(es) and Portuguese~(pt) is higher than the retrieval accuracy between. 
}

\ignore{
\paragraph{Linear Evaluation}
To evaluate the learned representations, we follow the widely used linear evaluation protocol \cite{chen2020simple}, where a linear classifier is trained in top of the frozen base network, and test accuracy is used as a proxy for representation quality.
In the fine-tuning stage, the batch size is 32. We use Adam optimizer with learning rate 7e-6 and warm-up steps 10000.
After each epoch, we test the fine-tuned model on the validation sets of all languages. 
We select the model with the best average result on the validation sets of all languages.

We follow the experiment setting in simCLR. Table 6 compares our results with previous approaches in the linear evaluation setting. we find Unicoder$_{uc}$ performs significantly better than XLM-R$_{base}$ and Unicoder$_{sa}$.
Especially, on these language without training data, Unicoder$_{uc}$ performs much better than Unicoder$_{sa}$ with more than 10 percentage,  which means Unicoder$_{uc}$ can learn high quality cross-lingual features by contrastive learning.

\begin{figure}[t]
\begin{center}
 \includegraphics[width=1.0\linewidth]{2021_acl_draft/news_vis.pdf}\vspace{-.3cm}
 \caption{Representation Visualization on News.}
 \label{fig-illustration}\vspace{-.3cm}
\end{center}
\end{figure}

\paragraph{Semi-supervised learning}
Since the pre-training of 108 language model is very time-consuming, we choose to search hyper-parameter on 2 language setting. Here we choose English and French as pre-training corpus. 
We follow the way in~\cite{chen2020simple} and sample 1\% or 10\% of the labeled training datasets in a class-balanced way.We simply fine-tune the whole base network on the labeled data without regularization. Our approach significantly
improves over state-of-the-art with both 1\% and 10\% of the labels.

In the fine-tuning stage, the batch size is 32. We use Adam Optimizer with learning rate 7e-6 and warm-up steps 10000.After each epoch, we test the fine-tuned model on the validation sets of all languages. 
We select the model with the best average result on the validation sets of all languages.

Table 7 shows the comparisons of our results against XLMR$_{base}$ and SOTA cross-lingual model Unicoder$_{sa}$ using parallel corpus. Our self-supervised model significantly outperforms  Unicoder$_{sa}$ on five languages in NC task.

}

\begin{table*}[!htp]
\small
    \begin{center}
          \resizebox{0.8\linewidth}{!}{
    	\begin{tabular}{cccccccccccccccccc}
        	\hline

          Methods & Metrics	& AR
 & BN & EN & FI & ID & JA & KO & RU & SW & TE & TH & AVG  \\
         	\hline

\multirow{2}{*}{MBERT} &
MPR@100 & 0.301 & 0.303 & 0.283 & 0.226 & 0.319 & 0.243 & 0.211 & 0.267 & 0.185 & 0.120 & 0.174  & 0.239\\
&Recall@100 & 0.695 & 0.712 & 0.749 & 0.645 & 0.739 & 0.662 & 0.565 & 0.674 & 0.537 & 0.433 & 0.529 & 0.631\\
        	\hline

\multirow{2}{*}{XLMR} &
MPR@100 & 0.365 & 0.374 & 0.275 & 0.318 & 0.395 & 0.299 & 0.304 & 0.306 & 0.274 & 0.346 & 0.401  & 0.333\\
&Recall@100 & 0.813 & 0.842 & 0.776 & 0.782 & 0.886 & 0.785 & 0.727 & 0.774 & 0.633 & 0.875 & 0.882 & 0.798\\
        	\hline

\multirow{2}{*}{InfoXLM} &
MPR@100 & 0.373 & 0.354 & 0.325 & 0.300 & 0.380 & 0.310 & 0.299 & 0.313 & 0.351 & 0.311 & 0.400  & 0.338\\
&Recall@100 & 0.806 & 0.860 & 0.804 & 0.749 & 0.869 & 0.788 & 0.717 & 0.767 & 0.724 & 0.867 & 0.874 & 0.802\\
        	\hline

\multirow{2}{*}{LABSE} &
MPR@100 & 0.372 & \textbf{0.504} & 0.314 & 0.309 & 0.376 & 0.271 & 0.309 & 0.325 & \textbf{0.394} & 0.465 & 0.374  & 0.365\\
&Recall@100 & 0.762 & \textbf{0.910} & 0.783 & 0.760 & 0.852 & 0.669 & 0.644 & 0.744 & 0.750 & \textbf{0.889} & 0.834 & 0.782\\
        	\hline

\multirow{2}{*}{CCP} &
MPR@100 & \textbf{0.426} & 0.457 & \textbf{0.359} & \textbf{0.372} & \textbf{0.462} & \textbf{0.377} & \textbf{0.346} & \textbf{0.360} & 0.392 & \textbf{0.470} & \textbf{0.489}  & \textbf{0.410}\\
&Recall@100 & \textbf{0.820} & 0.883 & \textbf{0.801} & \textbf{0.787} & \textbf{0.875} & \textbf{0.800} & \textbf{0.732} & \textbf{0.772} & \textbf{0.751} & 0.888 & \textbf{0.889} & \textbf{0.818}\\

        	\hline

        \end{tabular}
        }
        \end{center}
    \caption{Evaluation results on Mr. TYDI. We use MPR@100 and Recall@100 as evaluation metrics. Given each model, the corresponding retrieval results on all languages are listed in the same row.  }\label{tab:mytydi}
\end{table*}

\begin{table*}[!htp]
\tiny
    \begin{center}
          \resizebox{0.7\linewidth}{!}{
    	\begin{tabular}{cccccccccccccccccc}
        	\hline

          Methods & Metrics	& AR
 & BN & FI & JA & KO & RU & TE & AVG  \\
         	\hline






\multirow{2}{*}{XLMR} &
Recall@2kt & 0.414 & 0.470 & 0.529 & 0.407 & 0.439 & 0.566 & 0.639  & 0.452\\
&Recall@5kt & 0.534 & 0.572 & 0.611 & 0.498 & 0.540 & 0.397 & 0.718  & 0.553\\
        	\hline

\multirow{2}{*}{INFOXLM} &
Recall@2kt & 0.485 & 0.520 & 0.516 & 0.407 & \textbf{0.477} & 0.325 & 0.668  & 0.485\\
&Recall@5kt & 0.563 & 0.599 & 0.599 & 0.498 & 0.547 & 0.422 & 0.756 & 0.569\\
        	\hline

\multirow{2}{*}{LABSE} &
Recall@2kt & 0.469 & \textbf{0.553} & 0.487 & 0.382 & 0.418 & 0.333 & \textbf{0.676}  & 0.474\\
&Recall@5kt & \textbf{0.566} & \textbf{0.661} & 0.570 & 0.473 & 0.509 & \textbf{0.439} & \textbf{0.790} & \textbf{0.573}\\
        	\hline

\multirow{2}{*}{CCP} &
Recall@2kt & \textbf{0.472} & 0.493 & \textbf{0.570} & \textbf{0.423} & 0.456 & \textbf{0.342} & 0.672  & \textbf{0.490}\\
&Recall@5kt & 0.553 & 0.586 & \textbf{0.643} & \textbf{0.510} & \textbf{0.572} & 0.414 & 0.777 & 0.570\\
        	\hline


        \end{tabular}
        }
        \end{center}
    \caption{Evaluation results on XOR Retrieve. We use Recall@2kt and Recall@5kt as evaluation metrics. Given each model, the corresponding retrieval results on all languages are listed in the same row.  }\label{tab:xor}
\end{table*}

\subsection{Cross Lingual Query Passage Retrieval}

We further evaluate cross lingual transfer-ability on two new zero-shot settings: 1. Give a query from language $L$, retrieve relevant passages which can answer the query from the English corpus. 2. Give a query from language $L$, retrieve relevant passages which can answer the query from language $L$. 

Hence, we adopt XOR-QA~\cite{asai2020xor} dataset and Mr. TYDI~\cite{zhang2021mr} dataset to evaluate our method on the two settings. Both of the two datasets are constructed from TYDI, a question answering dataset covering eleven typologically diverse languages. The XOR-QA dataset consists of three tasks: XOR-Retrieve, XOR-English Span, and XOR-Full. We take the XOR-Retrieve task to evaluate our method. XOR-Retrieve is a cross-lingual retrieval task where the query is written in a target language (e.g., Japanese) and the model is required to retrieve English passages that can answer the query. Same to the source paper~\cite{asai2020xor}, we measure the recall by computing the fraction of the questions for which the minimal answer is contained in the top n tokens selected. We evaluate with n = 2k, 5k: R@2kt and R@5kt (kilo-tokens). The Mr. TYDI dataset is a multi-lingual benchmark dataset for mono-lingual query passage retrieval in eleven typologically diverse languages, designed to evaluate ranking with learned dense representations. Same to the source paper~\cite{zhang2021mr}, we use MRR@100 and Recall@100 as metrics.

In this paper, we adopt a zero-shot setting to evaluate our method. We train the pre-trained model on Natural Question data and directly test the model on XOR-QA~\cite{asai2020xor} dataset and Mr. TYDI~\cite{zhang2021mr} dataset.  We train the model on 8 NVIDIA Tesla V100 GPUs~(with 32GB RAM). We use AdamW Optimizer with a learning rate of 1e-5. The model is trained up to 20 epochs with a mini-batch size of 48. The rest hype-parameters are the same as DPR~\cite{karpukhin2020dense}. Besides, we also report our supervised result on XOR-Retrieve in appendix. We achieve 2\% advantage over the second-placed model. Besides, we also experiment our model on XOR Retrieve in supervised setting with the same hyperparameter settings, and  achieve SOTA results with 2\% advantage over the second-placed model on leaderboard\footnote{https://nlp.cs.washington.edu/xorqa/}.


In Table~\ref{tab:mytydi}, (1) we find CCP performs significantly better than XLM-R , InfoXLM and LaBSE. CCP achieves new SOTA results among methods without using bilingual data on the Mr. Tydi dataset. (2) Especially, we find InfoXLM has good performance on the 15 languages that it has bilingual corpus. However, on the languages that it doesn't have bilingual corpus, such as BN, FI, and TE, we find it's worse than LaBSE. For CCP, we find it has best performance on almost all languages, because our model is able to support 108 languages without collecting any bilingual corpus. In Table~\ref{tab:xor}, we find (1) CCP performs slightly better than InfoXLM and  have comparable performance on LaBSE. On these low-resource languages that InfoXLM can't cover, our method still have a better performance. (2) We find our model has especially bad performance on Bengali~(BN), which is consistent to results on Mr. TYDI.  Our performance is better than all models using bilingual data, which means our context-aware pretraining is very suitable to various retrieval tasks, not only bilingual paraphrase retrieval, also query passage retrieval.

\section{Ablation Study and Sensitivity Analysis}

\begin{table}[!h]
  \small
	\centering
	          \resizebox{0.8\linewidth}{!}{

	\begin{tabular}{ccccccccc}
		\hline
		\textbf{MB} &\textbf{$L_2$ norm} & \textbf{ABN} & \textbf{0.001} & \textbf{0.01} & \textbf{0.1} & \textbf{1.0}  \\
		\hline

        \checkmark& $\times$ & $\times$ & fail & fail & fail & fail  \\
		\checkmark&\checkmark & $\times$ & 70.3 &  71.6 & 72.5 & 63.1 \\
        \checkmark& $\times$ & \checkmark & 63.1 & 61.3 & 65.4 & 58.1  \\
        \checkmark& \checkmark & \checkmark & 85.7 & 84.8 & 91.3 & 75.3  \\
        $\times$ & \checkmark & \checkmark & 81.7 & 82.8 & 90.3 & 72.1  \\

        \hline
        \checkmark& \checkmark & BN & 58.1 & 60.1 & 62.4 & 61.5  \\
        
        \hline
	\end{tabular}
	}
	\caption{Examining the influence of $L_2$ normalization , batch normalization, temperature and memory bank with sentence retrieval task between En and Fr.}
	\label{tab:norm} 
\end{table}

\paragraph{Impacts of $L_2$ Normalization and Asymmetric Batch Normalization}

We next study the importance of $L_2$ normalization, batch normalization, and temperature $\tau$ in our contrastive loss. We use the pre-training and test setting in the last section. Table \ref{tab:norm} shows that without $L_2$ normalization before softmax and batch normalization in projection head, our model will fail in training. In the last row, we present the result of a model with $L_2$ normalization and vanilla batch normalization, and it has a very terrible result. The model appears to “cheat” the pretext task and easily finds a low-loss solution. This is possibly because the intra-batch communication among samples (caused by BN) leaks information.~\cite{he2020momentum}. With asymmetric batch normalization, different mean and variance will be used to calculate $z_c$ and $z_i$, respectively.

\ignore{
\begin{table}[!h]
  \small
	\centering

	\begin{tabular}{ccccccc}
		\hline
        \textbf{Model} & \textbf{EN2XX} & \textbf{XX2XX}   \\
		\hline
		Full & 92.2 & 85.1  \\
        w/o shifting & 89.4 & 84.2   \\
        w/o rotating & 91.5 & 84.7   \\
        w/o scaling & 91.7 & 84.9   \\
		\hline
	\end{tabular}
	
	\caption{Effect of three operations on cross-lingual sentence retrieval using accuracy measure. EN2XX means English-centric sentence retrieval, and XX2XX means non English-centric sentence retrieval.}
	\label{tab:ablation} 
\end{table}

\paragraph{Ablation Study of Cross-lingual Calibration}
Here we study the effectiveness of three operations in cross-lingual calibration on sentence retrieval task. We keep our approach with three operations as a reference. Then  we remove one operation at one time, and examine how it affects performance: (1) \underline{\emph{w/o shifting}} removes the shifting operation, (2) \underline{\emph{w/o rotating}} removes the rotation operation, and (3) \underline{\emph{w/o scaling}} removes the scale operation. The comparison results are reported in Table~\ref{tab:ablation}. Finally, we observe that the three kinds of operations contribute to the final performance. In particular, shifting operation seems to be more essential than the other two kinds of operations.

}

\begin{table}[!h]
  \small
	\centering
	\resizebox{0.8\linewidth}{!}{

	\begin{tabular}{ccccccccc}
		\hline
        \textbf{Windows Size /\ Batch Size} & \textbf{8} & \textbf{16} & \textbf{32} & \textbf{64}  \\
		\hline

		2 & 84.3 & 86.0 & 88.5 & 90.4 \\
        3 & 85.8 & 86.1 & 88.6 & 90.9  \\
        5 & 86.2 & 87.9 & 89.4 & 91.3  \\
		\hline
	\end{tabular}
	}
	\caption{Examining the influence of window size and batch size on the model performance between En and Fr of Tatoeba. Window size is 2 means predicting the next sentence.  }
	\label{tab:size} 
\end{table}


\paragraph{Impacts of Batch Size and Window Size}
Our current model is highly dependent on context window and batch size affects the number of negative samples directly. We study the importance of batch size and window size in the pre-training stage. Since 108 languages pre-training is too time-consuming, we only use English and French as pre-training corpus, and report the sentence retrieval result between English and French. Table \ref{tab:size} shows that without large window and large batch, performance is slightly worse.


%% file: 2022_naacl_draft/conclusion.tex
\section{Conclusion}
\label{sec:conclusion}
We propose a new cross-lingual pretrain task called Contrastive Context Prediction~(CCP), and conduct comprehensive evaluations with interesting findings observed. We find CCP task is able to make sentence embedding space of different language isomorphic.  The proposed approach achieves a excellent performance on multi-lingual dense retrieval. 